\documentclass[letterpaper]{article} 

\usepackage{aaai25}  
\usepackage{times}  
\usepackage{helvet}  
\usepackage{courier}  
\usepackage[hyphens]{url}  
\usepackage{graphicx} 
\usepackage{natbib}  
\usepackage{caption} 
\usepackage{amsmath}
\usepackage{float}
\usepackage{algorithm}
\usepackage{algorithmic}

\usepackage{newfloat}
\usepackage{listings}

\DeclareCaptionStyle{ruled}{labelfont=normalfont,labelsep=colon,strut=off} 
\floatstyle{ruled}
\newfloat{listing}{tb}{lst}{}
\floatname{listing}{Listing}
\title{Physics-Informed Neural Networks with Complementary Soft and Hard\\Constraints for Solving Complex Boundary Navier-Stokes Equations}
\author{
    Chuyu Zhou \equalcontrib\textsuperscript{\rm 1}, 
    Tianyu Li \equalcontrib\textsuperscript{\rm 2 }, 
    Chenxi Lan\textsuperscript{\rm 2}, 
    Rongyu Du\textsuperscript{\rm 4},
    Guoguo Xin\textsuperscript{\rm 1+}, 
    Pengyu Nan\textsuperscript{\rm 1},
    Hangzhou Yang\textsuperscript{\rm 1},
    Guoqing Wang\textsuperscript{\rm 2++},
    Xun Liu\textsuperscript{\rm 3},
    Wei Li \textsuperscript{\rm 3+++}
}
\affiliations{
    \textsuperscript{\rm 1}School of Physics, Northwest University, Xi'an 710127, China\\
     \textsuperscript{\rm 2}School of Computer Science and Engineering, University of Electronic Science and Technology of China, Chengdu 611731, China \\
     \textsuperscript{\rm 3}Department of Imaging Technology, Beijing Institute of Space Mechanics and Electricity, Beijing 100094, China
     \textsuperscript{\rm 4} School of Information Science \& Technology, Xi'an 710127, Northwest University
     \textsuperscript{\rm +}xinguo@nwu.edu.cn,\\
     \textsuperscript{\rm ++}qwang0420@hotmail.com,\\
     \textsuperscript{\rm +++}wei\_li\_bj@163.com
}

\usepackage{bibentry}

\begin{document}

\maketitle
\begin{abstract}
Soft- and hard-constrained Physics Informed Neural Networks (PINNs) have achieved great success in solving partial differential equations (PDEs). However, these methods still face great challenges when solving the Navier-Stokes equations (NSEs) with complex boundary conditions. To address these challenges, this paper introduces a novel complementary scheme combining soft and hard constraint PINN methods. The soft-constrained part is thus formulated to obtain the preliminary results with a lighter training burden, upon which refined results are then achieved using a more sophisticated hard-constrained mechanism with a primary network and a distance metric network. Specifically, the soft-constrained part focuses on boundary points, while the primary network emphasizes inner domain points, primarily through PDE loss. Additionally, the novel distance metric network is proposed to predict the power function of the distance from a point to the boundaries, which serves as the weighting factor for the first two components. This approach ensures accurate predictions for both boundary and inner domain areas. The effectiveness of the proposed method on the NSEs problem with complex boundary conditions is demonstrated by solving a 2D cylinder wake problem and a 2D blocked cavity flow with a segmented inlet problem, achieving significantly higher accuracy compared to traditional soft- and hard-constrained PINN approaches. Given PINN’s inherent advantages in solving the inverse and the large-scale problems, which are challenging for traditional computational fluid dynamics (CFD) methods, this approach holds promise for the inverse design of required flow fields by specifically-designed boundary conditions and the reconstruction of large-scale flow fields by adding a limited number of training input points. 

The code for our approach will be made publicly available.

\end{abstract}

\section{Introduction}

Fluid mechanics is an important field in science and engineering that deals with the study of the motion of liquids and gases. The Navier-Stokes equations are the basic partial differential equations that describe the dynamic behavior of viscous fluids, which are highly nonlinear partial differential equations and are widely used in aerodynamics, meteorology, oceanography, and industrial process simulations \cite{Munson_2006_04}. In most cases, its high degree of nonlinearity and complexity makes it difficult to obtain analytical solutions to equations, so almost study is based numerical solutions. Traditional numerical methods, such as finite difference methods (FDM), finite element methods (FEM), and finite volume methods (FVM), have made significant progress in solving the Navier-Stokes equations, commonly referred to as computational fluid dynamics (CFD) methods \cite{Anderson_1995}. However, these techniques need the sufficient initial and boundary conditions while they are often difficult to obtain in practice\cite{Abdel-Rahman_2011}. Even if accurate mathematical systems of NSEs are obtained, they still often encounter challenges with grid generation for the high-dimensional and the complex boundary condition problems, which require substantial computational resources and time, and may lead to issues of numerical instability and loss of accuracy \cite{Anderson_1995}.

Deep learning (DL) \cite{LeCun_2015_05} methods have achieved great success in computer science, particularly deep neural networks (DNNs), which possess a universal function approximation property that renders them promising as CFD surrogate models. The main achievements of deep learning methods have been in the area of image recognition, which relies on being trained on large datasets \cite{Krizhevsky_2017_05, He_2016_06}. In contrast, data from fluid systems are often sparse and noisy when obtained experimentally. Consequently, approaches that rely solely on data for learning, as in image recognition, are not viable for fluid models \cite{Eivazi_2024_04, Sharma_2023_02}. 

Physical Information Neural Network (PINNs) is a type of deep learning approach that incorporates physical constraints by embedding physical equations and boundary conditions into the loss function of a neural network. PINNs were first proposed by Raissi et al. in 2019 \cite{Raissi_2019_02} for solving one-dimensional partial differential equation problems, such as the Burgers equation, as well as inverse problems for two-dimensional and three-dimensional partial differential equations with a certain amount of labeled data \cite{Raissi_2019_02, Raissi_2020_01, Karniadakis_2021_05, Go_2023_11}. PINN offer significantly higher accuracy and efficiency compared to traditional CFD solvers when limited scattered partial spatio-temporal data are available for the flow problem under study. The structure of PINNs is inherently flexible, allowing the same formulation to be used for both forward and inverse problems. This eliminates the need for costly data assimilation techniques, which have historically slowed progress, especially in optimization and design tasks related to fluid dynamics \cite{Cai_2021_12}. Furthermore, PINN offer a unified approach to handling flow problems that exhibit phenomena across different scales \cite{Leung_2022_12}, for which  traditional CFD methods often require different models. However, for fluid models with complex boundary conditions, conventional PINN methods often struggle to accurately approximate both the boundary conditions and the partial differential equations \cite{Hsieh_2024_05}. Thus, the execution of boundary conditions is of paramount importance, and current methods for execute them in PINNs can be categorized into soft and hard constraints \cite{Barschkis_2023, Lu_2021_01}. Conventional PINNs typically adopt a soft-constraint approach \cite{Raissi_2019_02,Lai_2023}, where boundary conditions and initial conditions are explicitly embedded into the loss function, which is trained simultaneously with the PDE loss. By introducing the laws of physics into the loss function, the stability of the training process can be improved, and the network can avoid learning solutions that violate the laws of physics, especially when data is scarce. This can help models generalize better to unseen data, particularly when they need to extrapolate beyond the range of the training data. However, this method often fails to guarantee the satisfaction of both the PDE and the boundary conditions, resulting in lower accuracy of the outcomes \cite{Cuomo_2022_07, Bai_2022_11, Bischof_2021}. Hard-constraint methods \cite{Lu_2021_01} enforce boundary conditions by constructing a solution and a distance function that correspond to the boundary conditions, such that the network enforces the conditions during training by simply optimizing the partial differential equation loss, thereby avoiding the issue of tuning the weights of different parts of the loss function \cite{Lan_2023_05}. But the hard-constrained PINN approach encounters certain limitations. The solutions it devises for boundary conditions are often contingent upon a DNN with a limited number of hidden layers, which can hinder their ability to accurately solve complex boundary conditions. During training, this DNN solely uses coordinate points at the boundaries as input to ensure that boundary conditions are met. However, when boundary conditions are complex, the output of the DNN can become highly disordered, leading to extremely slow or non-convergent loss during the training of partial differential equations. Ultimately, this may prevent the model from producing accurate variable outputs. This limitation reduces the network's generative capabilities, especially in scenarios requiring higher flexibility or when dealing with intricate boundary conditions \cite{Barschkis_2023, Deng_2023}.

This paper constructs the solution of boundary conditions and the distance function specifically combine the soft- and the hard-constraint methods, enabling their application to various complex-boundary fluid dynamics problems. Its feasibility is demonstrated through testing on two fluid models under different boundary condition scenarios and comparing the results with those from CFD methods and traditional soft and hard constraint approaches.

In summary, our work makes the following contributions:
\begin{enumerate}
\item We propose a physics-informed neural network approach that combines soft and hard constraints to solve flow fields in regions with irregular obstructing structures.



\item A power function is proposed to redefine the distance metric in the hard-constrained approach for handling complex boundary conditions, thereby improving prediction accuracy.

\item The proposed method is thoroughly validated through extensive experimentation on complex-boundary Navier-Stokes equations, such as the 2D cylinder wake problem and the 2D blocked cavity flow problem with a segmented inlet.
\end{enumerate}

\section{Methodology}
In the this section, a brief background on the Navier-Stokes equations and the two boundary condition enforcement methods is discussed, followed by an explanation of the modification approach for the two constituent functions.

\subsection{Incompressible Navier-Stokes equations}
Incompressible Navier-Stokes equations capture a wide array of complex flow behaviors including vorticity, turbulence, and chaotic flows \cite{Tucker_2016, Majda_2002}. These equations are highly regarded for their ability to predict the conditions of flow in various scenarios. These equations are versatile and applicable to a broad range of problems in both engineering and science. They are used in the design of aircraft and automobiles, the study of blood flow in biomedical engineering, weather prediction, ocean currents, and many other areas. The incompressible Navier-Stokes equations are notoriously difficult to solve analytically except in the simplest cases. Most practical problems require numerical methods for their solution, which has driven substantial developments in CFD, numerical analysis, and computer science \cite{Zhang_2021_01}.

In the application of PINN, the form of the incompressible Navier-Stokes equations is generally given by \cite{Jin_2021_02}:

\begin{equation}
\begin{split}
 &\nabla \cdot \mathbf{u} = 0 \\
&\frac{\partial\mathbf{u}}{\partial t} + \left( \mathbf{u} \cdot \nabla \right)\mathbf{u} + \frac{1}{\rho}\nabla p - \nu\nabla^{2}\mathbf{u}-\mathbf{f} = 0 
\end{split}
\label{eq:NSEs}
\end{equation}
here, $\mathbf{u}$ is the velocity vector and, in two dimensions, it represents the velocity components $u$ in the x-direction and $v$ in the y-direction, which can be written in matrix form as $\mathbf{u} = (u, v)^T$. $p$ is pressure of the fluid. $\mathbf{u}$ and $p$ are typically functions of space and time, and they are also the physical quantities that we aim to solve for in Eq. \ref{eq:NSEs}. $\rho$ and $\nu$ are density and dynamic viscosity coefficient, respectively, and, for incompressible flows, they are characteristic parameters of the fluid and generally do not vary with space and time. $\mathbf{f}$ is the external force exerted in the fluid. For the flow field model considered in this article, the external force $f$ is neglected.

\subsection{Soft constraints}

In soft-constraint PINN methods, the boundary conditions (BC) and initial conditions (IC) are typically expressed in the loss function as \cite{Raissi_2019_02}:

\begin{equation}
\mathcal{L}_\text{BC} = \frac{1}{N_\text{BC}}\sum_{i = 1}^{N_\text{BC}}\left\| \mathbf{u}(x_i,t_i) - \mathbf{u}_\text{BC}(x_i,t_i) \right\|^{2},
\label{Eq:sofBound}
\end{equation}
and
\begin{equation}
    \mathcal{L}_\text{IC} = \frac{1}{N_\text{IC}}\sum_{i = 1}^{N_\text{IC}}\left\| \mathbf{u}(x_i,t_0) - \mathbf{u}_\text{IC}(x_i,t_0) \right\|^{2}.
\label{Eq:sofIni}
\end{equation}

The expression for the loss of NSEs is given by:

\begin{equation}
\begin{split}
\mathcal{L}_\text{PDE} = &\frac{1}{N_\text{PDE}}\sum_{i = 1}^{N_\text{PDE}} \left( \left\| \nabla \cdot \mathbf{u} \right\|^{2} + 
\right.\\ &\left.
\left\| \frac{\partial\mathbf{u}}{\partial t} + \left(\mathbf{u} \cdot \nabla\right)\mathbf{u} + \frac{1}{\rho}\nabla p - \nu\nabla^{2}\mathbf{u} \right\|^{2} \right)
\end{split}
\label{Eq:softLoss}
\end{equation}

The loss function of an unregularized PINN is given by:

\begin{equation}
    \mathcal{L =}\lambda_{1}\mathcal{L}_\text{PDE} + \lambda_{2}\mathcal{L}_\text{IC} + \lambda_{3}\mathcal{L}_\text{BC},
\label{Eq:softWLoss}
\end{equation}
where $\lambda_i$ indicate the loss weights for corresponding items. By adjusting the loss weights, the network can be biased during the training process, thereby enhancing the accuracy of the output. However, tuning these weights can be complex, and a single joint loss function may not ensure that all conditions are satisfied, particularly when dealing with intricate boundary conditions. To avoid simultaneously optimizing the condition loss and PDE loss, a more rigorous approach, known as the hard-constraint method, has been proposed.

\begin{figure*}[t]
\centering
\includegraphics[width= 2.0\columnwidth]{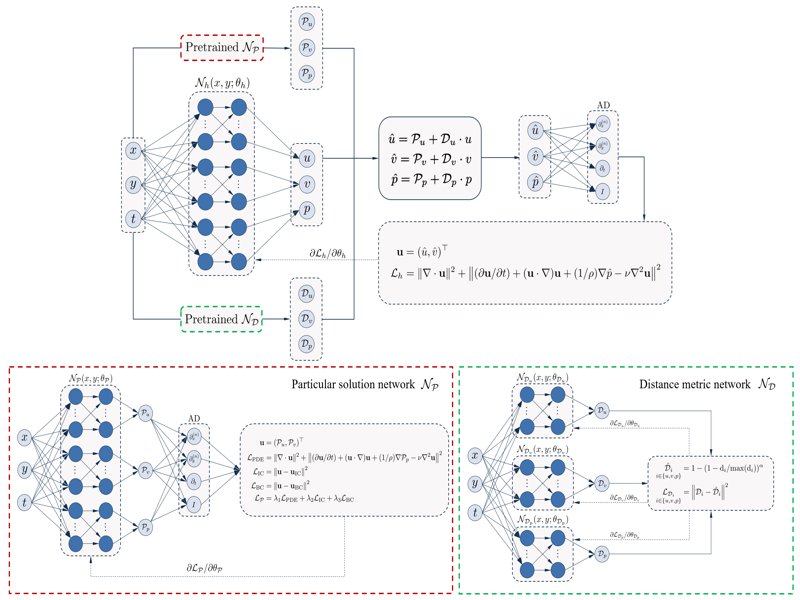} 
\caption{The architecture of the proposed method.}
\label{Fig:figFrame}
\end{figure*}

\subsection{Hard constraints}

The main idea behind the hard-constraint method is to construct an additional solution function that precisely captures the boundary conditions, along with a smooth function that represents the distance from the spatial coordinates to the boundary. By utilizing these two constructed functions, the network's output can be modified during training, thereby enforcing the prescribed boundary conditions without the need for simultaneous training with the PDE loss.

In the fluid model, modifications to the network output, i.e., the velocity $\hat{\mathbf{u}}$ and pressure $\hat{p}$, take the form \cite{Lu_2021_01}
\begin{equation*}
    \hat{\mathbf{u}} = \mathcal{P}_{\mathbf{u}}(x,t) + \mathcal{D}_{\mathbf{u}}(x,t) \cdot \mathbf{u}(x,t)
\end{equation*}
and
\begin{equation*}
    \hat{p} = \mathcal{P}_{p}(x,t) + \mathcal{D}_{p}(x,t) \cdot p(x,t)
\end{equation*}
where, $\mathcal{P}_{\mathbf{u}}$ and $\mathcal{P}_{p}$ denote the additional solution functions that satisfy the boundary conditions only. $\mathcal{D}_{\mathbf{u}}$ and $\mathcal{D}_{p}$ are the distance functions with a value of $0$ at the boundary. The modified partial differential equation loss is expressed as

\begin{align*}
    \mathcal{L}_\text{PDE} =& \frac{1}{N_\text{PDE}}\sum_{i = 1}^{N_\text{PDE}}\left( \left\| \nabla \cdot \hat{u} \right\|^{2} +\right. \\
    & \left.\left\| \frac{\partial \hat{u}}{\partial t} + \left( \hat{\mathbf{u}} \cdot \nabla \right) \hat{\mathbf{u}} + \frac{1}{\rho}\nabla \hat{p} - \nu\nabla^{2}\hat{\mathbf{u}} \right\|^{2} \right)
\end{align*}
This expression is of the same form as the soft constraint form Eq.(\ref{Eq:softLoss}), except that $\mathbf{u}$ and $p$ are replaced by $\hat{\mathbf{u}}$ and $\hat{p}$, respectively. In this form, the enforcement of the boundary conditions specified by $\mathcal{P}_{\mathbf{u}}$ at the boundary is guaranteed since the distance function evaluates to zero at the boundary.
In the case of strong-constraint methods, where the boundary conditions and geometric models are relatively simple, $\mathcal{P}_{\mathbf{u}}$ and $\mathcal{D}_\mathbf{u}$ can be expressed analytically, but in most cases, analytical expressions cannot be constructed. Therefore, the solution can be obtained through three separate DNNs, which are the particular solution network $\mathcal{N}_{P}$ representing IC/BC, the distance metric network $\mathcal{N}_{D}$, and the primary network $\mathcal{N}_{h}$ satisfying the solution of the NSEs. The networks $\mathcal{N}_{P}$ and $\mathcal{N}_{D}$ are pretrained, so the final construction of the solution in the two-dimensional incompressible flow model is as follows:

\begin{align*}
&\mathcal{N}_q = \mathcal{N}_{\mathcal{P}_q}(x,t) + \mathcal{N}_{\mathcal{D}_q}(x,t) \cdot \mathcal{N}_{h_q}(x,t),
\end{align*}
here, q represents u, v, and p.

The loss function of the particular solution network $\mathcal{N}_{\mathcal{P}_q}$ is constructed by:

\begin{equation}
\begin{split}
    \mathcal{L}_{\mathcal{P}_q} = \frac{1}{N_\text{BC}}\sum_{i = 1}^{N_\text{BC}}\left\| \mathcal{Q}_{\mathcal{P}} - \mathcal{Q}_{B} \right\|^{2} \\
    + \frac{1}{N_\text{IC}}\sum_{i = 1}^{N_{IC}}\left\| \mathcal{Q}_{\mathcal{P}} - \mathcal{Q}_{I} \right\|^{2},
\end{split}
\label{Eq:lossLpu}
\end{equation}
where $N_\text{BC}$ and $N_\text{IC}$ denote the number of training points taken at the boundary and initial locations, respectively, while $\mathcal{Q}_{\mathcal{P}}$ represents the network's output of $q$. $\mathcal{Q}_{B}$ and $\mathcal{Q}_{I}$ represent the prescribed values for the boundary and initial conditions of $q$, respectively. It is noteworthy that in traditional hard-constraint methods, $\mathcal{N}_p$ is treated as a mere network for boundary conditions, training them separately from the governing equations, without considering their physical impact inside the solution domain. 

The loss function of the distance metric network $\mathcal{N}_{\mathcal{D}_q}$ is constructed as follows:
\begin{equation}
\begin{split}
    \mathcal{L}_{\mathcal{D}_q} = \frac{1}{N_{\mathcal{D}_q}}\sum_{i = 1}^{N_{\mathcal{D}_q}}\left\| \mathcal{D}_q - \hat{\mathcal{D}}_q \right\|^{2},
\end{split}
\label{Eq:distLoss}
\end{equation}
here, $N_{\mathcal{D}_q}$ is the number of training points taken within the spatial-temporal domain, $\mathcal{D}_q$ is the network output, and $\hat{\mathcal{D}}_q$ is the distance from a training point to the spatial-temporal boundary, as shown below:
\begin{equation*}
\hat{\mathcal{D}}_{q} =\min(\text{distance\ to\ the\ spatiotemporal\ boundary\ of\ } q), 
\end{equation*}
Typically, the $\mathcal{N}_{\mathcal{P}_q}$ network is only trained on the boundaries without regularization in the spatial domain. For two-dimensional incompressible flows with regular boundary conditions, especially at the inlet, the primary network $\mathcal{N}_{h_q}$ can converge the loss while enforcing the boundary conditions. However, since the $\mathcal{N}_{\mathcal{P}_q}$ network is not constrained in the interior, its outputs within the inner spatial domain might be irregular and deviate from the requirements of the partial differential equations, leading to slow convergence of the loss function during the training of the primary network. Moreover, training only the boundary can even lead to the failure of the primary network’s loss function to converge when the boundary conditions are complex. Therefore, a tailored construction of the $\mathcal{N}_{\mathcal{P}_q}$ network is required for models with more intricate boundary conditions.

\subsection{Modifications}

In our approach, as shown in Fig. \ref{Fig:figFrame}, we modify the particular solution network $\mathcal{N}_{\mathcal{P}_q}$ in the hard-constraint method by substituting its output with that of a softly-constrained network to avoid unphysical effects of the boundary conditions on the inner solution region. In the softly-constrained network, the weight associated with the boundary loss is increased, allowing for simultaneous training of the boundary conditions and rudimentary training of the space in the vicinity of the boundaries with respect to the partial differential equation. The advantage of doing so is that one can leverage the improved hard-constraint approach for training regardless of how intricate the boundary conditions (in particular, the initial conditions) are, while also speeding up convergence of the master network’s loss function.

\begin{figure}[t]
\centering
\includegraphics[width= 0.6\columnwidth]{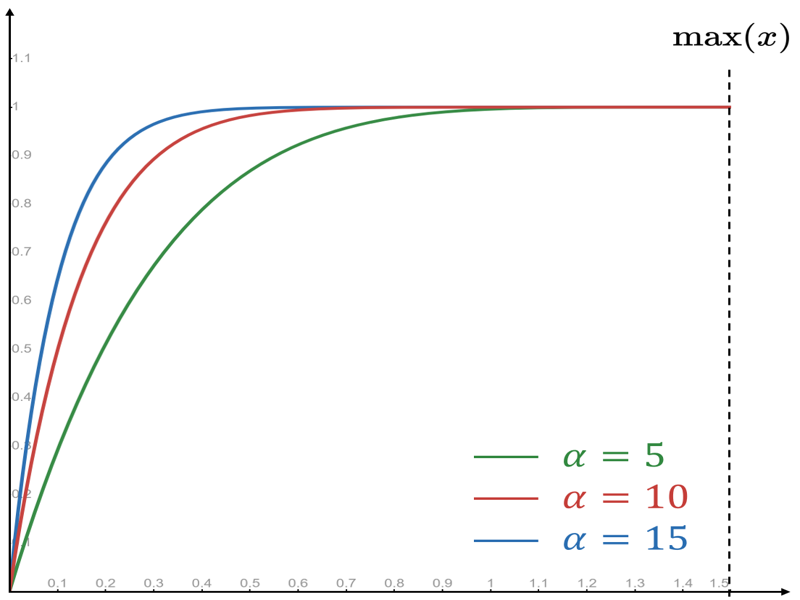} 
\caption{The impact of different $\alpha$ values on the the power function $f_p$.}
\label{Fig:figPower}
\end{figure}

We also make a simple modification to the distance metric network $\mathcal{N}_{\mathcal{D}_q}$, using a power function of the distance $\hat{\mathcal{D}_q}$ instead of the original distance function as the label for training, so that the output of the network approaches $0$ near the boundary and quickly rises to $1$ away from it, while still being dominated by the training results of the main network $\mathcal{N}_{h_q}$ in the spatial interior. The formula of the power function $f_p$ is given by:

\begin{equation}
\begin{split}
f_q(\hat{D}_q)=1-(1-\hat{D}_q/max(\hat{D}_q))^\alpha,
\end{split}
\label{Eq:powerfunc}
\end{equation}
where parameter $\alpha$ is a positive value used to control the rate of growth. Then the loss function of the distance metric network is changed from eq. (\ref{Eq:distLoss}) to

\begin{equation}
\begin{split}
    \mathcal{L}_{\mathcal{D}_q} = \frac{1}{N_{\mathcal{D}_q}}\sum_{i = 1}^{N_{\mathcal{D}_q}}\left\| \mathcal{D}_q - f_q\left(\hat{\mathcal{D}}_q\right) \right\|^{2},
\end{split}
\label{Eq:ndistLoss}
\end{equation}
Fig. \ref{Fig:figPower} illustrates the relationship between $\hat{D}_q$ and $f_q(\hat{D}_q)$ when $\alpha$ is 5, 10, and 15 respectively. As can be seen, when a point is near the boundary, the curve becomes steeper, whereas it becomes smoother for points farther from the boundary. A large value of $\alpha$ encourages the distance metric network to focus more on points close to the boundary while being less sensitive to those further away, thereby reducing the network’s requirements. However, if $\alpha$ is too large, it causes the network to predict a distance value approaching 1 even when a point is very close to the boundary, which incorrectly diminishes the impact of the particular solution network on these points. In our experiment, $\alpha$ is set to 10 by trial and error.

The architecture of our approach is shown in Fig. \ref{Fig:figFrame}. The overall structure consists of three sub-networks: $\mathcal{N}_{\mathcal{P}_q}$, $\mathcal{N}_{\mathcal{D}_q}$, and $\mathcal{N}_{h_q}$, representing the particular solution network, distance metric network, and primary network, respectively. Among them, $\mathcal{N}_{P}$ and $\mathcal{N}_{D}$ are pre-trained networks, and only the parameters of the $\mathcal{N}_{h}$ network are optimized during final training.

The particular solution network $\mathcal{N}_{P}$ is composed of a regular soft-constrained PINN structure, with loss function defined in eq. (\ref{Eq:softWLoss}). To ensure accurate enforcement of boundary conditions, the weight of the boundary loss is adjusted to a large value. The distance metric network $\mathcal{N}_{D}$ is a low-layer DNN network trained using power functions of distance as the standard values. The primary network (for training equations) $\mathcal{N}_{h}$ follows the same format as the original hard-constrained method, but only trains the modified output variables $u$, $v$, and $p$ to satisfy the partial differential equations.

\section{Experiments}
To better demonstrate the advantages of our approach and to allow readers to verify the results with relatively small computational effort, both cases in the experimental section are based on two-dimensional steady incompressible flows, hence the consideration of initial conditions in Eq. (\ref{Eq:sofIni}), the loss term of the initial condition in Eq. (\ref{Eq:lossLpu}), and the time derivative in the NSEs described by Eq. (\ref{eq:NSEs}) are omitted.
In this section, we first introduce the experimental setting. We then discuss two case studies: the two-dimensional flow around a cylinder and the segmented inlet with obstructed square cavity flow. Taking CFD calculation results as the ground truth (GT), we compare our approach (mPINN) with traditional soft-constrained PINN (sPINN) and hard-constrained PINN (hPINN) methods to demonstrate the advantages of our approach in solving complex boundary NSEs. Additionally, to assess the impact of our two strategies-modifying the particular solution network (mP) with a soft-constrained condition and the distance metric network (mD) with a power function—we present the results from mP and mD as part of the ablation analysis.

\subsection{Experimental setting}
Three separate DNNs are employed to individually train the three variables of the NSEs, i.e., $\mathcal{N}_{\mathcal{P}}$, $\mathcal{N}_{\mathcal{D}}$, and $\mathcal{N}_{h}$, with architectures as shown in Fig. \ref{Fig:figFrame}. In the three networks, two-dimensional spatial coordinates serve as inputs, while the output includes the velocity $u$, $v$, and $p$. For the networks $\mathcal{N}_{\mathcal{P}}$ and $\mathcal{N}_{\mathcal{D}}$, a DNN structure with four hidden layers, each containing 20 neurons, is adopted. For the network $\mathcal{N}_{h}$, a DNN structure with six hidden layers, each having 120 neurons, is employed. The activation function is chosen to be the $\tanh$ function, and the Adam algorithm \cite{Kingma_2014} is utilized as the optimizer. A temperature annealing learning rate strategy \cite{Wang_2021_01} is implemented, with specific adjustments made according to the model. Build the network architecture using the machine learning platform PyTorch and accelerate iterations on a graphics processing unit (GPU).

\begin{figure}[h]
\centering
\includegraphics[width=0.8\columnwidth]{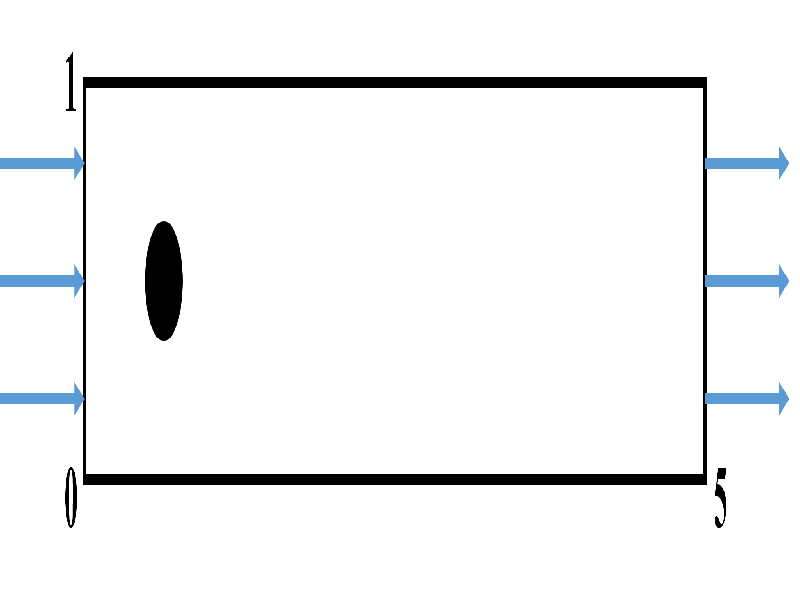}
\caption{Scheme of the two-dimensional flow around a cylinder.}
\label{Fig:figCylinder}
\end{figure}

\begin{figure}[h]
\centering
\includegraphics[width=0.8\columnwidth]{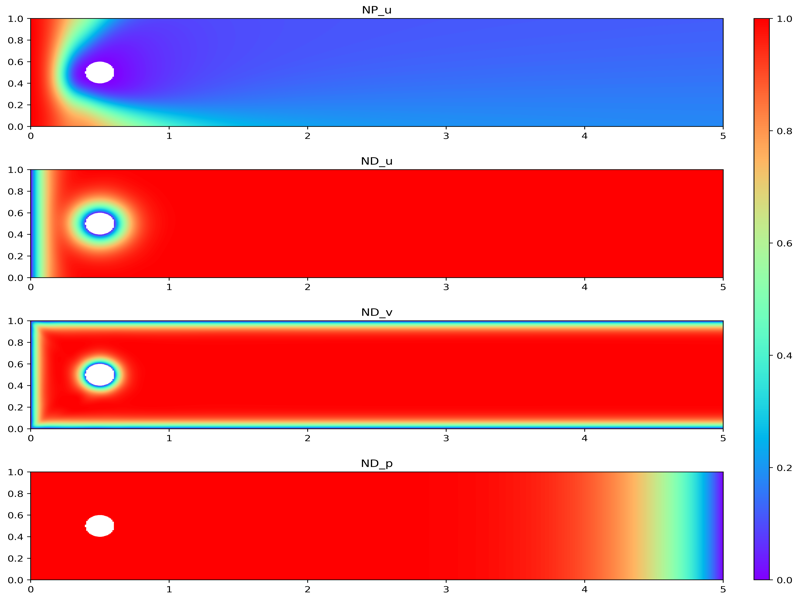} 
\caption{The trained boundary prediction for Case 1. NP\_u represents the result of the particular solution network for u, while ND\_u, ND\_v, and ND\_p respectively represent the results of the distance metric network for u, v, and p}
\label{Fig:figCBD}
\end{figure}

\begin{figure}[h]
\centering
\includegraphics[width=1.0\columnwidth]{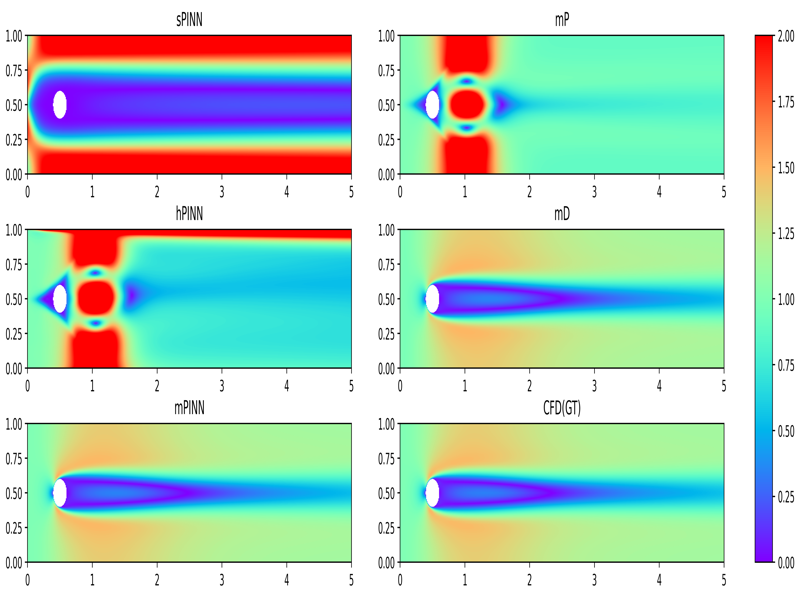} 
\caption{Velocity distributions for the flow around a cylinder from different methods.}
\label{Fig:figPredC}
\end{figure}

\subsection{Case 1: Two-dimensional flow around a cylinder}
The cylindrical flow around model is a classic model in fluid dynamics, the specific model is set as shown in Fig.\ref{Fig:figCylinder}, the pipe length is set to 5, the width is set to 1, and there is a circular obstruction with $(0.5, 0.5)$ as the center of the circle, the inlet boundary condition sets the normal inflow velocity $u$ as 1, the outlet boundary condition sets the static pressure as 0, and the normal component of velocity on the rest of the boundary is set to 0. The boundary conditions are:

\begin{equation*}
    \begin{cases}
u = 1,& x = 0,0 \leq y \leq 1 \\
u \cdot n_x = 0,& \text{others} 
\end{cases} 
\end{equation*}
\begin{equation*}
    v \cdot n_y = 0,\quad \text{ on  }\partial\Omega
\end{equation*}
\begin{equation*}
    p = 0,\quad x = 5,0 \, \leq y \leq 1
\end{equation*}
The soft-constraint PINN loss function adopted for this model takes the form:
\begin{equation*}
\begin{split}
    \mathcal{L}_{\mathcal{P}} =&  \lambda_{1}\Big( \left\| \nabla \cdot \hat{u} \right\|^{2} + \left\| \partial\hat{u} / \partial t + \left(\hat{u} \cdot \nabla \right)\hat{u} + 1 / \rho \nabla\hat{p} - \right.\\ &\left.
    \nu\nabla^{2}\hat{u} \right\|^{2}_{x \in \Omega} \Big)  + \lambda_{2}\left( \left\| \mathcal{U}_{\mathcal{P}} - \mathcal{U}_{B} \right\|_{x \in \partial\Omega}^{2}
    \right. \\ &\left.
    + \left\| p_{\mathcal{P}} - p_{B} \right\|_{x \in \partial\Omega}^{2} \right)
\end{split}
\end{equation*}

Here, $\Omega$ denotes the interior of the geometric region, while $\partial\Omega$ denotes the boundary. The loss weight $\lambda_1$ for the equation part is set to 1, and the loss weight $\lambda_2$ for the boundary part is set to 10, with the aim of making the network focus more on the training of boundary conditions.

The network $\mathcal{N}_{\mathcal{D}}$ is also trained using three separate DNNs for $u$, $v$, and $p$, each with four hidden layers of 20 neurons. Training is performed using the Adam optimizer for 10,000 iterations with a learning rate of $10^{-3}$. The loss function takes the form of eq. (\ref{Eq:ndistLoss}).

 The outputs of the $\mathcal{N}_{\mathcal{P}}$ and $\mathcal{N}_{\mathcal{D}}$ networks are shown in Fig.\ref{Fig:figCBD}.

The final modification to the variable result is as follows:
\begin{equation*}
    \hat{u} = N_{\mathcal{P}_u}(x,t) + \mathcal{N}_{\mathcal{D}_u}(x,t) \cdot N_{h_u}(x,t)
\end{equation*}
\begin{equation*}
    \hat{v} = \mathcal{N}_{\mathcal{D}_v}(x,t) \cdot N_{h_v}(x,t)
\end{equation*}
\begin{equation*}
    \hat{p} = \mathcal{N}_{\mathcal{D}_p}(x,t) \cdot N_{h_p}(x,t)
\end{equation*}

Since both $v$ and $p$ are set to 0 at the boundary, there is no need to configure the boundary results separately; they are simply constrained by the distance function.

\begin{figure}[h]
\centering
\includegraphics[width=0.7\columnwidth]{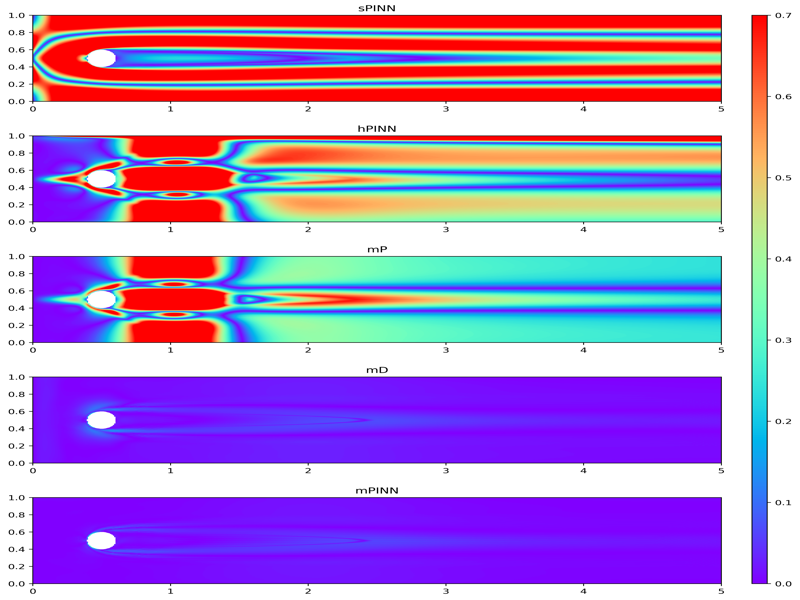} 
\caption{The residuals of sPINN, hPINN, mP, mD, and our mPINN compared to the GT in Case 1.}
\label{Fig:figErrorC}
\end{figure}

In the case of 2D cylindrical pipe flow model, the qualitative results from sPINN, hPINN, mP, mD, mPINN are shown in Fig.\ref{Fig:figPredC}. The residuals of these methods compared with CFD results are shown in Fig.\ref{Fig:figErrorC}. 

\begin{figure}[H]
\centering
\includegraphics[width=0.4\columnwidth]{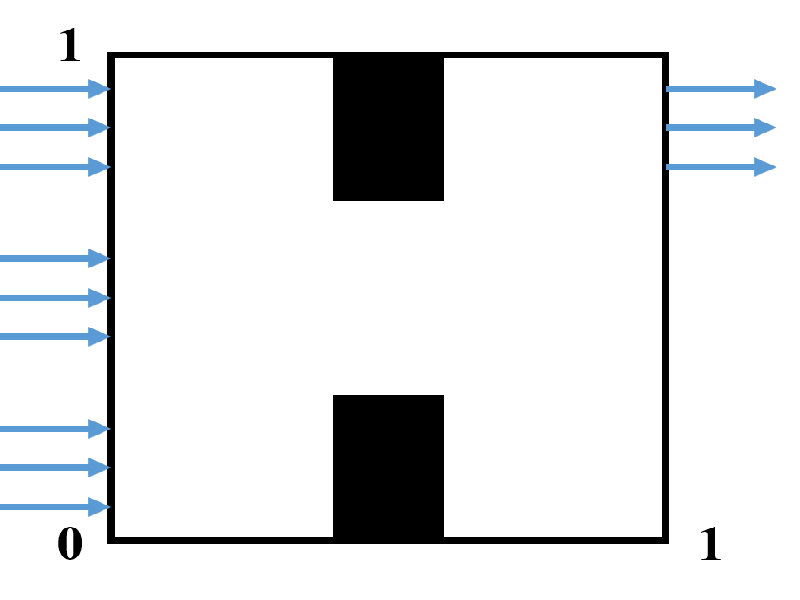}
\caption{Scheme of the Segmented inlet with obstructed square cavity flow.}
\label{Fig:figSquare}
\end{figure}

\subsection{Case 2: Segmented inlet with obstructed square cavity flow}
The model is set with two rectangular obstructions on the top and bottom of the square cavity. The inlet is segmented, as shown in Fig. \ref{Fig:figSquare}. The square cavity has a side length of 1, with the rectangular obstructions having a width of 0.2 and a height of 0.3. The boundary conditions are set such that the normal inflow velocity $u$ at the inlet is 0.5, the static pressure at the outlet is 0, and the normal component of velocity on all other boundaries are set to 0. The boundary conditions are:

\begin{equation*}
\begin{cases}
u = 0.5,& x = 0,\ 0 \leq y \leq 0.2, \\
&0.4 \leq y \leq 0.6, 0.8 \leq y \leq 1 \\
u \cdot n_x = 0,& \text{others}
\end{cases} 
\end{equation*}
\begin{equation*}
    v \cdot n_y = 0,\quad \text{in } \partial\Omega
\end{equation*}
\begin{equation*}
    p = 0,\quad  x = 1, 0.8 \leq y \leq 1
\end{equation*}

\begin{figure}[h]
\centering
\includegraphics[width=0.7\columnwidth]{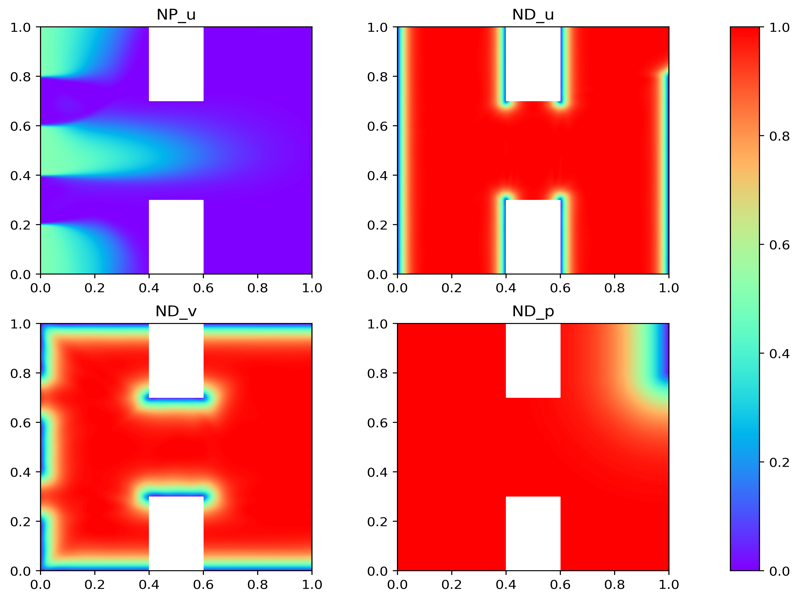}.
\caption{The trained boundary prediction for Case 2. NP\_u represents the result of the particular solution network for u, while ND\_u, ND\_v, and ND\_p respectively represent the results of the distance metric network for u, v, and p}
\label{Fig:figSBD}
\end{figure}

The model is simulated using the modified strong enforcement method, The $\mathcal{N}_{\mathcal{P}}$, $\mathcal{N}_{\mathcal{D}}$, and $\mathcal{N}_h$ networks employ DNNs with 6 hidden layers of 100 neurons each, 4 hidden layers of 20 neurons each, and 10 hidden layers of 120 neurons each, respectively. The output results of the boundary condition solution function network related to $u$ and the distance function network for $u$, $v$, and $p$ are shown in Fig.\ref{Fig:figSBD}.

\begin{figure}[h]
\centering
\includegraphics[width=0.9\columnwidth]{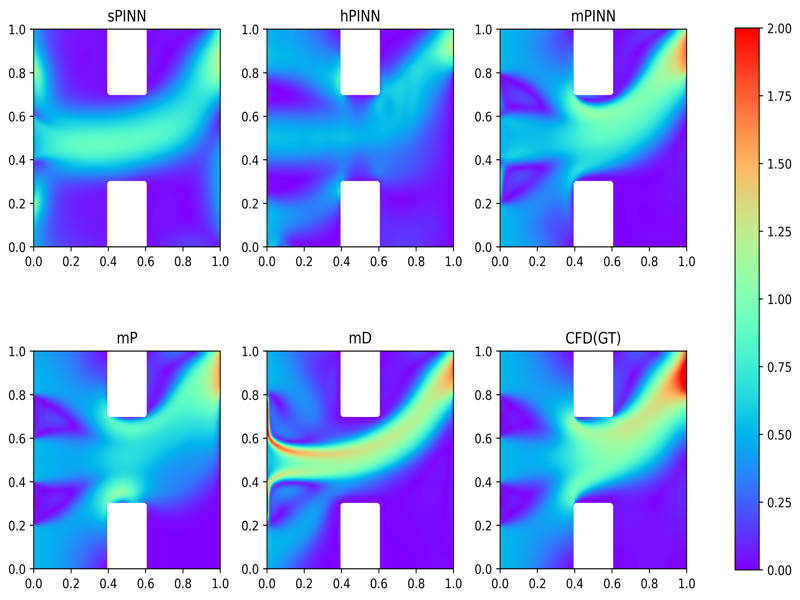}.
\caption{Comparison of velocity distributions for Case 2 across different methods.}
\label{Fig:figPredS}
\end{figure}

\begin{figure}[h]
\centering
\includegraphics[width= 1.0\columnwidth]{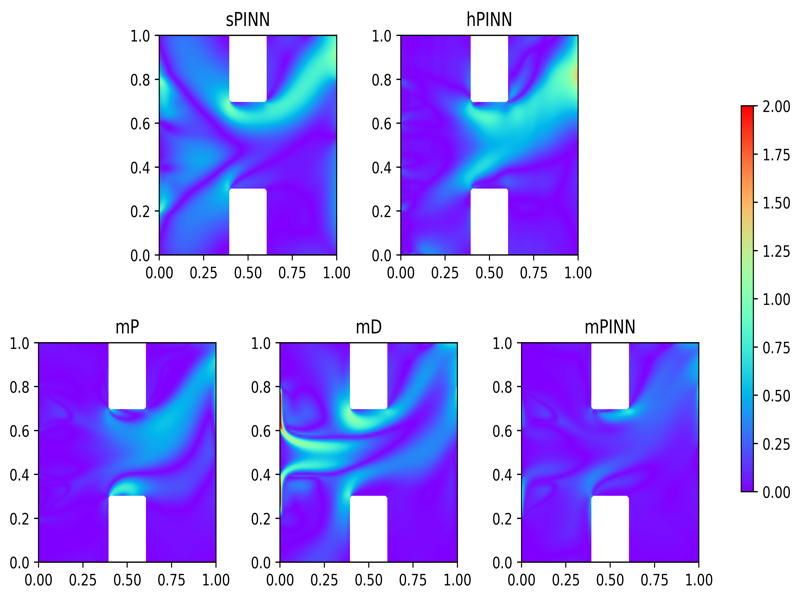} 
\caption{The residuals of sPINN, hPINN, mP, mD, and our mPINN compared to the GT in Case 2.}
\label{Fig:figErrorS}
\end{figure}

The velocity distributions for case 2 from sPINN, hPINN, mP, mD, and our mPINN are shown in Fig.\ref{Fig:figPredS}. The original hard-constraint method fails to simulate the segmental inlet boundary condition, with the equation loss in the $\mathcal{N}_{h}$ network failing to converge. The modified hard-constraint method is demonstrated to be capable of simulating the model under complex inlet boundary conditions when compared with the results of the soft-constraint method. The residuals of these methods compared with CFD results are shown in Fig. \ref{Fig:figErrorS}.

\subsection{Result analysis}

The quantitative results of the mean square error between CFD calculation and the predictions from sPINN, hPINN, mP, mD and mPINN for the two cases are presented in Table \ref{Tab:tab}. Combined with the qualitative results shown in Fig.\ref{Fig:figPredC},\ref{Fig:figErrorC},\ref{Fig:figPredS},\ref{Fig:figErrorS}, we can conclude that our mPINN can accurately solve complex boundary flow problems with interior barriers, while the traditional soft-constrained PINN (sPINN) method and hard-constrained PINN (hPINN) method incur significant errors in solving such problems. 
Moreover, both strategies—enforcing soft boundary conditions (mP) and optimizing the distance function (mD)—can improve prediction results, demonstrating the effectiveness of these approaches. However, we also observe that the effectiveness of the two strategies varies with different boundary conditions, as shown in Table \ref{Tab:tab}, mD performs better in Case 1, while mP is more effective in Case 2. Using both methods simultaneously to achieve the best results further validates the robustness of our approach.

\begin{table}[h]  
\centering  
\begin{tabular}{c|cc|cc|c}
\hline
 &sPINN& hPINN& mP & mD &mPINN \\ \hline \hline 
Case 1 & 0.4958 & 0.5023 & 0.2610 & 0.0006& 0.0003\\ \hline 
Case 2 & 0.0925 & 0.0889 & 0.0335 & 0.0691& 0.0184 \\ \hline  
\end{tabular}  
\caption{The mean square errors of sPINN, hPINN, mP, mD, and our mPINN relative to the CFD results in the two cases.}
\label{Tab:tab}
\end{table}

\section{Conclusion}
PINNs hold great potential as surrogate models for fluid dynamics; however, their application is often hindered by the complexity of the NSEs, compounded by intricate boundary conditions. This paper improves the traditional hard-constraint approach by pretraining the particular solution network using a soft-constraint PINN method and optimizing the distance metric network with a power function. These modifications make the approach applicable to a broader range of complex fluid models and accelerate the convergence of the loss function. Through testing on two-dimensional steady incompressible flows with three different boundary conditions, our method yields the highest accuracy, serving as the benchmark data for CFD computations, as compared to traditional soft-constraint and hard-constraint PINN approaches. We also tested additional cases provided in the supplementary materials, which demonstrate that the introduced method effectively handles general NSEs with complex boundary conditions.

\bibliography{aaai25}

\end{document}